# Towards Accountable AI in Eye Disease Diagnosis: Workflow, External Validation, and Development

Qingyu Chen, PhD[1,2,*,+], Tiarnan D L Keenan, BM BCh, PhD[3,*], Elvira Agron, MA[3], Alexis Allot, PhD[1], Emily Guan, BS[1], Bryant Duong, MS[1], Amr Elsawy, PhD[1], Benjamin Hou, PhD[1], Cancan Xue, MS[4], Sanjeeb Bhandari, MD[3], Geoffrey Broadhead, MD[5], Chantal Cousineau-Krieger, MD[3], Ellen Davis, MD[6], William G Gensheimer, MD[7,8], Cyrus A Golshani, MD[9], David Grasic, MD[10], Seema Gupta, MD[6], Luis Haddock, MD[11], Eleni Konstantinou, MD[3], Tania Lamba, MD[12], Michele Maiberger, MD[9], Dimosthenis Mantopoulos, MD[8], Mitul C Mehta, MD[13], Ayman G Elnahry, MD[3], Mutaz AL-Nawaflh, MD[3], Arnold Oshinsky, MD[9], Brittany E Powell, MD[14], Boonkit Purt, MD[15], Soo Shin, MD[9], Hillary Stiefel, MD[6], Alisa T Thavikulwat, MD[3], Keith James Wroblewski, MD[16], Tham Yih Chung, MD, PhD[4,17,18], Chui Ming Gemmy Cheung, MD, PhD[4,17,18], Ching-Yu Cheng, MD, PhD[4,17,18], Emily Y Chew, MD[3], Michelle R. Hribar, PhD[3,6], Michael F. Chiang, MD[3], Zhiyong Lu, PhD[1,+]

1. National Library of Medicine, National Institutes of Health, Maryland, USA
2. Yale School of Medicine, Yale University, New Haven, USA
3. National Eye Institute, National Institutes of Health, Bethesda, USA
4. Yong Loo Lin School of Medicine, National University of Singapore, Singapore
5. Save Sight Institute, Sydney University, Sydney, Australia
6. Casey Eye Institute, Oregon Health & Science University, Oregon, USA
7. VA Healthcare System, Vermont, USA
8. Dartmouth Hitchcock Medical Center, Lebanon, New Hampshire, USA
9. VA Healthcare System, Washington DC, USA
10. Carolina Vision Center, Fayetteville, USA
11. Bascom Palmer Eye Institute, University of Miami, Florida, USA
12. Krieger Eye Institute, Baltimore, USA
13. Gavin Herbert Eye Institute, University of California, Irvine, USA
14. Fort Belvoir Community Hospital, Virginia, USA
15. Uniformed Services University of the Health Sciences, Bethesda, USA
16. George Washington University Hospital, The George Washington University, Washington, DC, USA
17. Singapore National Eye Center, Singapore
18. Duke-NUS Medical School, National University of Singapore, Singapore

*These authors contributed equally to this work.

+Corresponding authors:

Qingyu Chen, PhD

Department of Biomedical Informatics and Data Science

Yale School of Medicine

Yale University

qingyu.chen@yale.edu

Zhiyong Lu, PhD

Division of Intramural Research

National Library of Medicine (NLM)

National Institutes of Health (NIH)

zhiyong.lu@nih.gov

**KEY POINTS**

Question:

What are the gaps in downstream accountability of medical AI, focusing on AI-assisted workflows, external validation, and further development?

Findings:

In a multi-round, multi-institutional study of 24 clinicians grading 2,880 AMD risk features from 240 patients, AI assistance improved diagnostic accuracy by up to 50%, enhanced time efficiency, and had higher generalization through further development. Yet, clinicians with AI assistance did not always yield the highest performance, especially compared with using AI alone, underscoring challenges in explainability and clinician trust.

Meaning:

Systematic evaluations of AI workflows, external validation, and further development are essential for closing accountability gaps and ensuring effective, generalizable medical AI.

## Abstract

**Importance**

Timely disease diagnosis is challenging due to limited clinical availability and growing burdens. Although artificial intelligence (AI) shows expert-level diagnostic accuracy, a lack of downstream accountability— including workflow integration, external validation, and further development— continues to hinder its real-world adoption.

**Objective**

To address gaps in the downstream accountability of medical AI through a case study on age-related macular degeneration (AMD) diagnosis and severity classification.

**Design, Setting, and Participants**

We developed and evaluated an AI-assisted diagnostic and classification workflow for AMD. Four rounds of diagnostic assessments (accuracy and time) were conducted with 24 clinicians from 12 institutions. Each round was randomized and alternated between Manual and Manual + AI, with a washout period. In total, 2,880 AMD risk features were evaluated across 960 images from 240 Age-Related Eye Disease Study patient samples, both with and without AI assistance. For further development, we enhanced the original DeepSeeNet model into DeepSeeNet+ using ~40,000 additional images from the US population and tested it on three datasets, including an external set from Singapore.

**Main Outcomes and Measures**

We measured the F1-score for accuracy (Wilcoxon rank-sum test) and diagnostic time (linear mixed-effects model), comparing Manual vs. Manual + AI. For further development, the F1-score (Wilcoxon rank-sum) was again used.

**Results**

Among the 240 patients (mean age, 68.5 years; 53% female), AI assistance improved accuracy for 23 of 24 clinicians, increasing the average F1-score by 20% (37.71 to 45.52), with some improvements exceeding 50%. Manual diagnosis initially took an estimated 39.8 seconds per patient, whereas Manual + AI saved 10.3 seconds and remained 1.7–3.3 seconds faster in later rounds. However, combining manual and AI may not always yield the highest accuracy or efficiency, underscoring challenges in explainability and trust. DeepSeeNet+ performed better in three test sets, achieving 13% higher F1-score in the Singapore cohort.

**Conclusions and Relevance**

In this diagnostic study, AI assistance improved both accuracy and time efficiency for AMD diagnosis. Further development was essential for enhancing AI generalizability across diverse populations. These findings highlight the need for downstream accountability during early-stage clinical evaluations of medical AI. All code and models are publicly available.

# Introduction

Timely disease diagnosis remains challenging due to rising disease burdens, limited clinician availability, and insufficient health care access[1-3]. Although artificial intelligence (AI) has demonstrated potentials[3-6], many medical AI studies conclude after reporting testing-set metrics alone, neglecting how AI integrates into clinical workflows, external validation for diverse populations, and continual algorithmic refinement[7-11].

Our work addresses these gaps through a case study of age-related macular degeneration (AMD), a leading cause of vision loss[12-16]. We developed an AI-assisted diagnostic workflow and conducted four rounds of assessments—measuring both accuracy and efficiency—with 24 clinicians, comparing performance with and without AI assistance. We also refined the AI algorithm using ~40,000 additional medical images and tested it on three benchmarks, including an external Singapore cohort. Overall, our study underscores the importance of downstream accountability during early-stage clinical evaluations of medical AI, moving beyond reliance on internal benchmarking test sets alone. The code and models are publicly available at the repository[17].

# Materials and Methods

### The clinical standard of AMD diagnosis

The current standard of care for diagnosing AMD in clinical practice and performing prognostic stratification for risk of progression to late AMD is the Age-Related Eye Diseases Study (AREDS) Simplified Severity Scale[18]. The diagnostic/severity classification procedure comprises two steps: First, three macular risk features are identified and categorized (Individual risk factors) separately for both eyes: drusen (ranging from 0 to 2), pigmentary abnormalities (ranging from 0 to 1), and late AMD (ranging from 0 to 1). Second, these individual risk feature scores for both eyes are used to calculate the overall AMD severity level, for that individual, on a scale of 0 to 5. The quantification and interpretation of each severity level is summarized in eTable 1 in Supplementary 1.

### Clinician participants and the AI model

The AI-assisted diagnostic workflow is designed to support clinicians in diagnosing and classifying the severity of AMD. This study conducts multi-round and head-to-head comparisons involving AI, and 24 clinicians from 12 institutions to evaluate both diagnostic accuracy and time efficiency with and without AI assistance.

DeepSeeNet[5] was selected as the AI model to assist in the diagnosis and severity classification of AMD within the workflow, owing to its state-of-the-art performance and free availability. This model was trained and validated on ~ 60,000 images from a longitudinal study of about 4,500 individuals, representing a broad range of AMD severity—from none to advanced disease.

Twenty-four clinicians from 12 different institutions were recruited to participate in the evaluation of AI-assisted diagnostic/severity classification workflow. The clinicians included 13 retina specialists and 11 ophthalmologists who were not retina specialists (i.e., general ophthalmology and other subspecialties), as detailed in Supplementary 2.

### Design and implementation of the AI-assisted diagnostic workflow

Figure 1 illustrates the AI-assisted diagnostic workflow. To assess the effectiveness of the AI-assisted AMD diagnosis and severity classification workflow, we conducted a comprehensive comparative analysis. Institutional review board approval was obtained and written informed consent for the research was obtained from all study participants.

We first created an evaluation dataset from the test set of the DeepSeeNet model, as shown in Figure 1(A). The gold standard labels originate from expert human grading of the images at the Wisconsin Reading Center, as described in detail previously[5,19].

In brief, two human experts graded the images for the AMD risk features, then a computerized algorithm calculated the severity levels. In the case of any discrepancy regarding the severity level between the graders, a senior investigator would adjudicate the final severity level. To create the evaluation dataset, we randomly selected 40 samples for each AMD severity level, ranging from 0 to 5, i.e., to generate an equally distributed dataset. This resulted in a total of 240 patient samples comprising 480 color fundus photographs. Subsequently, we randomly divided these samples into four separate batches and ensured that images from each patient were present in only one batch, as shown in Figure 1(B).

The comparative analysis was completed in four rounds, as shown in Figure 1(C). Each round comprised two batches: one batch where clinicians annotated images manually (e.g., Batch A in Figure 1(C)), referred to as **Manual**, and another batch where they annotated with AI assistance (e.g., Batch B in Figure 1(C)), referred to as **Manual + AI**. Clinicians provided their final diagnosis/severity scores in both scenarios, while their diagnostic times were tracked. The order of Manual versus Manual + AI was reversed in each pair of rounds (e.g., Rounds 1 vs. 2 and Rounds 3 vs. 4 in Figure 1(C)).

After the first two rounds, each clinician had graded all samples exactly once—half under Manual and half under Manual + AI. At this point, a washout period of a month was introduced: the batches were renamed (A→E, etc.), the samples within each batch were randomly reordered, and their IDs were changed to prevent memorization and bias. The same clinicians then repeated the process in another two rounds, with batches that had been presented as Manual now presented as Manual + AI, and vice versa (e.g., Batch A, initially Manual, became Batch E under Manual + AI in Figure 1(C)). Additionally, the order of presentation was reversed (e.g., if Round 1 started with Manual, Round 3 started with Manual + AI in Figure 1(C)). By the end, each clinician had graded all samples exactly twice—once under Manual and once under Manual + AI.

**Evaluation measures**

This study was conducted and reported in accordance with the Standards for Reporting Diagnostic Accuracy Studies guidelines. The evaluation assesses whether AI assistance could enhance both the **effectiveness and time efficiency of diagnostic/severity classification**. Effectiveness was quantified using the F1-score as the primary metric, complemented by precision, specificity, and sensitivity. The statistical analysis is detailed in **Statistical Analysis (Quantifying AI-assisted effectiveness)**. Efficiency was measured by the time taken in seconds per patient for diagnosis. The statistical analysis is detailed in **Statistical Analysis (Quantifying AI-assisted efficiency)**.

**Further development and external validations**

We further curated an additional 39,916 fundus images from 2,940 participants with AMD in the Age-Related Eye Diseases Study 2 (AREDS2) and split these into 70%, 10%, and 20% for training, validation, and testing sets, respectively. The AREDS2 was a multicenter phase 3 randomized controlled clinical trial designed to enroll individuals at moderate to high risk of progression to late AMD[20,21]. We combined the two training sets (i.e., from AREDS and AREDS2) to perform additional training on the model, then compared the performance of the **New Model (named DeepSeeNet+)** with the **Original Model (DeepSeeNet)** on three test sets:

(1) the 240 patient samples from the original AREDS testing set, described above, referred to as **the AREDS set**.
(2) 150 patient samples with a simplified AREDS scale level of 3 to 5 (50 patients per level), focused on intermediate and late AMD, from the testing set of the AREDS2 dataset, referred to as **the AREDS2 set**.
(3) 180 patient samples with a simplified AREDS scale level of 1 to 5 from the Singapore Epidemiology of Eye Diseases (SEED) Study[22], referred to as **the SEED set**.

The SEED Study is a multi-ethnic, population-based study aimed at providing insights into the epidemiology of eye diseases across three major ethnic groups (Malay, Indian, and Chinese) in Singapore. We used this set to evaluate the performance of the models on external populations. The statistical analysis is detailed in **Statistical Analysis (Quantifying external validation performance)**. The training detail is also provided in Supplementary 3.

**Statistical Analysis**

Quantifying AI-Assisted effectiveness. A two-sided Wilcoxon rank-sum test (p-value of <0.05) was used to compare the F1-scores for Manual and Manual + AI. All analyses were conducted in Python using the SciPy package.

Quantifying AI-Assisted time efficiency. Each clinician participated in four rounds, annotating images both manually and with AI assistance while their diagnostic times were recorded. To minimize confounding factors, we alternated the order of manual versus AI-assisted tasks in each round. Nevertheless, within-clinician correlation and practice effects may still occur. To address these, we employed a linear mixed-effects model (p-value of <0.05) with random intercepts for each clinician, treating round (1–4), method (Manual vs. Manual + AI), and their interaction as fixed effects. The analysis was conducted in Python (v 3.13.2) using the statsmodels (v0.14.4) package.

Quantifying external validation performance. For each of three test sets, we performed bootstrapping with a random sample size of 60 (drawn without replacement) over 100 iterations for both DeepSeeNet and DeepSeeNet+. We then applied a two-sided Wilcoxon rank-sum test (p-value of <0.05) to compare their F1-scores. As above, the analysis was carried out in Python using the SciPy (v1.15.2) package.

## Results

### Patient information

From the AREDS study, we selected 240 patients by randomly choosing 40 samples for each AMD severity level (0–5). The patients had a mean age of 68.5 years, and 53% were female.

**Missing data**

As described above, a four-round AI-assisted diagnostic workflow was conducted with 24 clinician participants from 12 institutions. All 24 clinicians' annotations were properly captured in each round. However, diagnostic times were only recorded across all four rounds for 19 of those 24 clinicians, as five clinicians had missing data in one of the rounds.

**Incorporating AI into the clinical workflow**

Figure 2 details the diagnostic and classification performance, comparing F1-scores for both Manual and Manual + AI scenarios. Figure 2(A) illustrates the overall AMD severity level. Overall, the results indicate an improvement in manual AMD diagnostic/classification accuracy with the AI assistance in 23 out of the 24 clinicians. Specifically, the average F1-score for grading AMD severity increased by more than 20%, rising from 37.71 (95% CI, 0.2783–0.4417) to 45.52 (95% CI, 0.3901–0.5161; P-value < 0.0001). Secondary evaluation metrics also show Manual + AI consistently outperformed Manual (see eTable 2 in Supplementary 4).

For drusen size grading Figure 2(B), the average F1-score increased with AI assistance by ~11%, from 59.83 (95% CI, 0.4925–0.6845) to 66.22 (95% CI, 0.5954- 0.7069; P-value <0.0001), with 23 out of 24 clinicians demonstrating higher diagnostic accuracy when using AI assistance. Similarly, for pigmentary abnormality grading Figure 2(C), the average F1-score increased by ~10%, from 68.11 to 75.00 (P-value <0.0001), with 23 out of 24 clinicians demonstrating higher diagnostic accuracy with AI assistance. Finally, for late AMD grading Figure 2(D), the average F1-score increased by ~8%, from 56.98 (95% CI, 0.3282–0.7262) to 61.26 (95% CI, 0.4511- 0.7430; P-value 0.18), with 15 out of 24 clinicians demonstrating higher diagnostic accuracy with AI assistance.

Figure 2(E) shows the diagnostic performance changes of the overall AMD severity scale at the individual clinician grader level. It demonstrates that the F1-score of individual clinicians on the overall AMD severity scale improved with AI assistance over a range of 5 to 50%. 23 demonstrated higher diagnostic accuracy when using AI assistance. There was only one instance where the performance of Manual + AI did not improve, but the difference was only 0.2% lower, with the F1-score changing from 42.77 to 42.51.

Figure 3 provides more detailed performance results for Manual vs. Manual + AI at each severity level in the overall AMD severity level (Figure 3 (A)) and its three constituent risk features (Figure 3 (B)-(D)). The results show that manual accuracy was improved with AI assistance consistently at each severity level. Notably, for the overall AMD severity level, the largest improvement was observed in the final AMD severity scale of 3, where the F1-score was enhanced by almost 50% with AI assistance, from 23.09 to 34.36. This was largely due to the improved accuracy of drusen and pigment grading with AI assistance. The F1 score of medium drusen and large drusen improved by 22% from 33.94 to 41.17, and 8% from 71.90 to 77.89, respectively. Similarly, the F1 score of the presence of pigment improved by 10% from 68.11 to 75.00. In contrast, for grading late AMD presence, while its F1-score improved by 7% from 56.98 to 61.26 with the AI assistance, performance was still relatively lower than that of other two risk factors.

The diagnostic time was recorded properly in all four rounds for 19 of the 24 clinician participants. Figure 4(A) illustrates the mean diagnostic times for Manual and Manual + AI across these rounds, while Figure 4(B) provides a more granular, physician-level breakdown. Detailed results are available in

Supplementary 5. In Round 1, Manual required an estimated 39.8 seconds (95% CI, 34.1–45.6); AI assistance reduced this by 10.3 seconds (p < 0.001). Although Manual times improved by approximately 12–14 seconds in Rounds 2–4 (indicating a learning curve or practice effect), AI assistance remained significantly faster by 1.7–3.3 seconds in each subsequent round (all p < 0.05). Overall, these findings suggest that AI assistance provides a meaningful reduction in diagnostic time, even after accounting for practice effects over repeated testing. Note that other factors may also affect these results. For instance, the random intercept variance was 108.3, indicating substantial baseline differences in clinicians' speeds.

**Further development**

Table 1 shows the performance on three external testing sets of both models. The results demonstrate that DeepSeeNet +had superior performance: for the AREDS2 set and the SEED set, its F1-score increased by 25% and 29%, respectively, and had superior performance at almost all severity level. For the AREDS set that was used in the evaluation of AI-assisted diagnostic workflow (the same distribution of the training set from the original AI model), the F1-score remained the same.

## Discussions

**Primary findings**

First, AI assistance improved both manual diagnostic/classification accuracy and time efficiency. Specifically, the results indicate that AI assistance enhanced grading performance for overall AMD severity as well as individual AMD risk features, among both retina specialists and comprehensive ophthalmologists. In terms of efficiency, AI assistance significantly reduced diagnostic times in the first round, with the benefit decreasing but remaining evident in later rounds after accounting for practice effects over repeated testing. Collectively, these findings demonstrate the potential of integrating AI assistance into clinical practice.

Second, while Manual + AI often outperformed Manual alone, it did not necessarily exceed the performance of AI alone. As shown in Figure 2, for overall AMD severity, AI alone produced higher F1 scores than Manual + AI in 19 of 24 cases. For drusen grading, AI alone had higher F1 scores in 17 of 24 cases, and, for pigmentary abnormality grading, AI alone exceeded Manual + AI in 19 of 24 cases. In contrast, for late AMD grading, 18 of the 24 clinicians achieved higher performance than AI alone. This finding highlights the ongoing challenges in making medical AI algorithms explainable and trustworthy for clinicians.

Third, a systematic comparison between DeepSeeNet and DeepSeeNet+ revealed that the latter had superior performance, particularly on external populations. This underscores the importance of continued development and refinement of existing medical AI algorithms.

**Implications for downstream accountability of AI in healthcare**

Previous studies have highlighted the importance of downstream accountability in AI-assisted eye disease diagnosis, citing limited external validations, minimal evaluations with ophthalmologists, and insufficient ongoing monitoring[7-11]. For instance, one study reported that AI-assisted diabetic retinopathy diagnosis had a positive predictive value of only 12% in 193 patients during external validation[23]. Although multiple studies have compared AI performance directly with ophthalmologists[5,6,24,25], very few have assessed whether AI assistance actually improves clinicians' diagnostic accuracy or time efficiency.

In particular, the Collaborative Community on Ophthalmic Imaging Working Group pointed out that benchmarking the performance of a combined human and AI model workflow is understudied[26].

Beyond ophthalmology, downstream accountability of AI in healthcare has consistently been a concern[27-30]. As mentioned above, a comprehensive review of 82 studies on AI disease classification algorithms across 18 disease specialties (e.g., cardiology, respiratory, and ophthalmology) stressed that only 14 studies compared performance with that of healthcare professionals, and only 25 studies included external validation[7]. Similar observations are also reported in other reviews[8-11]. Importantly, of the 14 studies, none evaluated how AI might assist healthcare professionals in improving diagnostic accuracy and time efficiency—an essential goal, since AI disease classification algorithms are primarily intended to support clinical expertise.

Studies further show that there is little evidence for improved clinician performance with AI assistance[31,32]. Such issues have been highlighted by the Developmental and Exploratory Clinical Investigations of DEcision support systems driven by Artificial Intelligence (DECIDE-AI), which provides expert consensus statements on the reporting guidelines for early-stage clinical evaluation[33]. Our study helps fill this gap in downstream accountability by showing precisely how AI assistance can enhance manual diagnosis and be integrated into clinical workflows. Notably, while AI assistance holds promise for improving diagnostic workflows, challenges—such as ensuring the highest possible accuracy and efficiency through Manual + AI—remain. These findings illustrate why early-stage clinical evaluations of medical AI must extend beyond internal benchmarking test sets and include thorough, real-world assessments of how AI assists clinicians.

## Limitations

Our study has several limitations. First, the evaluation of the AI-assisted disease diagnosis workflow uses samples from the test set of the original AI model, as mentioned in the Methods section. This may lead to an overestimation of AI performance. Nevertheless, as shown in Table 1, its overall performance on three independent datasets, including the test set, remains similar. In the future, we aim to use data from new patients with consent for further evaluations.

Second, although we have implemented a stringent evaluation study design and minimized potential biases, AMD may still be challenging to grade based on color fundus photographs alone, as previously mentioned. This could result in lower overall diagnostic accuracy. We will explore additional imaging modalities and tools to enhance performance in clinical scenarios.

## Conclusions

In this diagnostic study involving 24 clinicians from 12 institutions, diagnosing 2,880 AMD risk features in 240 patients across multiple rounds (with and without AI assistance), the results demonstrated that AI assistance improved both accuracy and time efficiency. Following additional training, the AI model also showed broader generalizability across diverse populations. Nonetheless, challenges such as algorithm explainability and clinician trust remain. The results illustrated the need for downstream accountability during early-stage clinical evaluations of medical AI—beyond relying solely on internal benchmarking—and highlighted the importance of ongoing validation and refinement to ensure broader applicability in clinical practice and ultimately facilitate more effective disease management.

## Data sharing statement

The AREDS dataset is available at dbGaP[34]. Investigators wishing to gain access should submit an online Data Access Request. Anonymized data are available at the individual participant level. The codes and models are publicly available at the repository[17].

## Acknowledgement

This research was supported by R00LM014024 and the NIH Intramural Research Program of the National Library of Medicine and the National Eye Institute. Dr. Mehta also acknowledges support from a departmental unrestricted grant by Research to Prevent Blindness. The views expressed in this study do not reflect the official policy or position of the U.S. Army Medical Department, the U.S. Army Office of the Surgeon General, the Department of the Army, the Department of Defense, the Uniformed Services University of the Health Sciences, or any other agency of the U.S. Government. The funding agencies had no role in the design or conduct of the study; data collection, management, analysis, or interpretation; manuscript preparation, review, or approval; or the decision to submit the manuscript for publication.

The authors Q.C. and T.K. had full access to all the data in the study and take responsibility for the integrity and accuracy of the data and analyses.

# Tables

Table 1. F1-scores for diagnosing AMD using the DeepSeeNet and DeepSeeNet+ models across three datasets: AREDS, AREDS2, and SEED. Scores are detailed for each severity scale (0-5) with averages, comparing performance between the two models.

|  |  | Final AMD Scale | DeepSeeNet model | DeepSeeNet+ model | P-value |
|---|---|---|---|---|---|
| **F1-Score per Scale** | **AREDS** | Overall | 0.4755 (0.3066 - 0.6434) | 0.4793 (0.3091 - 0.6554) | 0.95 |
|  |  | 0 | 0.6852 | 0.6667 |  |
|  |  | 1 | 0.3704 | 0.3797 |  |
|  |  | 2 | 0.2821 | 0.2927 |  |
|  |  | 3 | 0.4390 | 0.3421 |  |
|  |  | 4 | 0.5882 | 0.5833 |  |
|  |  | 5 | 0.6349 | 0.7302 |  |
|  | **AREDS2** | Overall | 0.5162 (0.3738 - 0.6447) | 0.6395 (0.4873 - 0.7947) | < .001 |
|  |  | 3 | 0.4211 | 0.4872 |  |
|  |  | 4 | 0.4091 | 0.6491 |  |
|  |  | 5 | 0.7391 | 0.8163 |  |
|  | **SEED** | Overall | 0.3895 (0.3050 - 0.4745) | 0.5243 (0.4438 - 0.6100) | < .001 |
|  |  | 0 | 0.5915 | 0.6275 |  |
|  |  | 1 | 0.3125 | 0.5000 |  |
|  |  | 2 | 0.2609 | 0.1923 |  |
|  |  | 3 | 0.3396 | 0.4478 |  |
|  |  | 4 | 0.3158 | 0.7077 |  |
|  |  | 5 | 0.5538 | 0.7385 |  |

# Figures

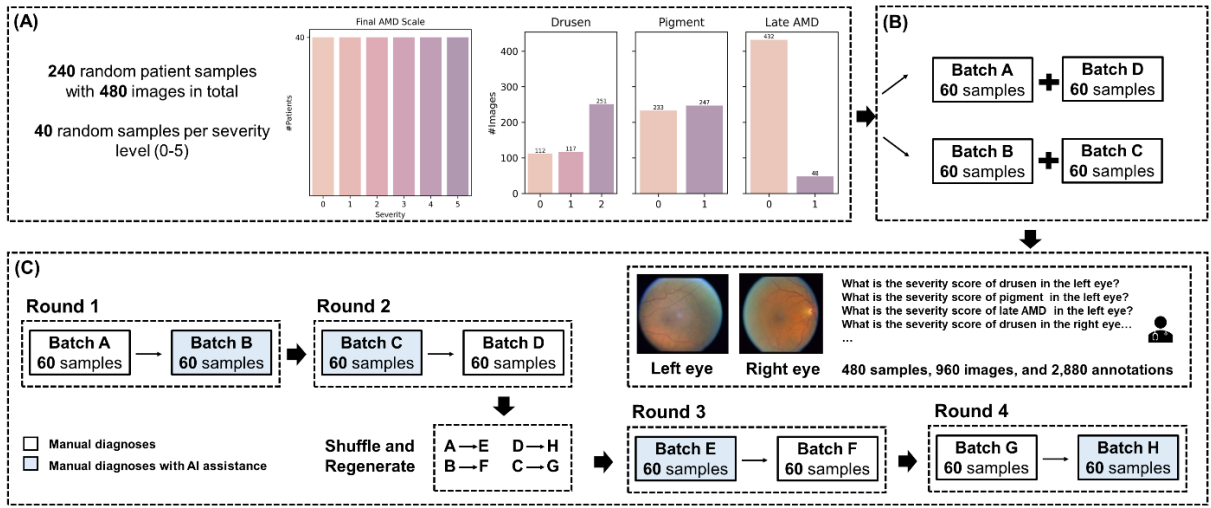

Figure 1. Overview of the AI-assisted diagnostic/classification workflow. (A) Dataset of 240 patients with 480 images, 40 samples per severity level, with distribution of risk factors—drusen, pigment abnormalities, and late AMD. (B) Division of samples into batches. (C) Evaluation pipeline where clinicians grade color fundus photographs with and without AI assistance.

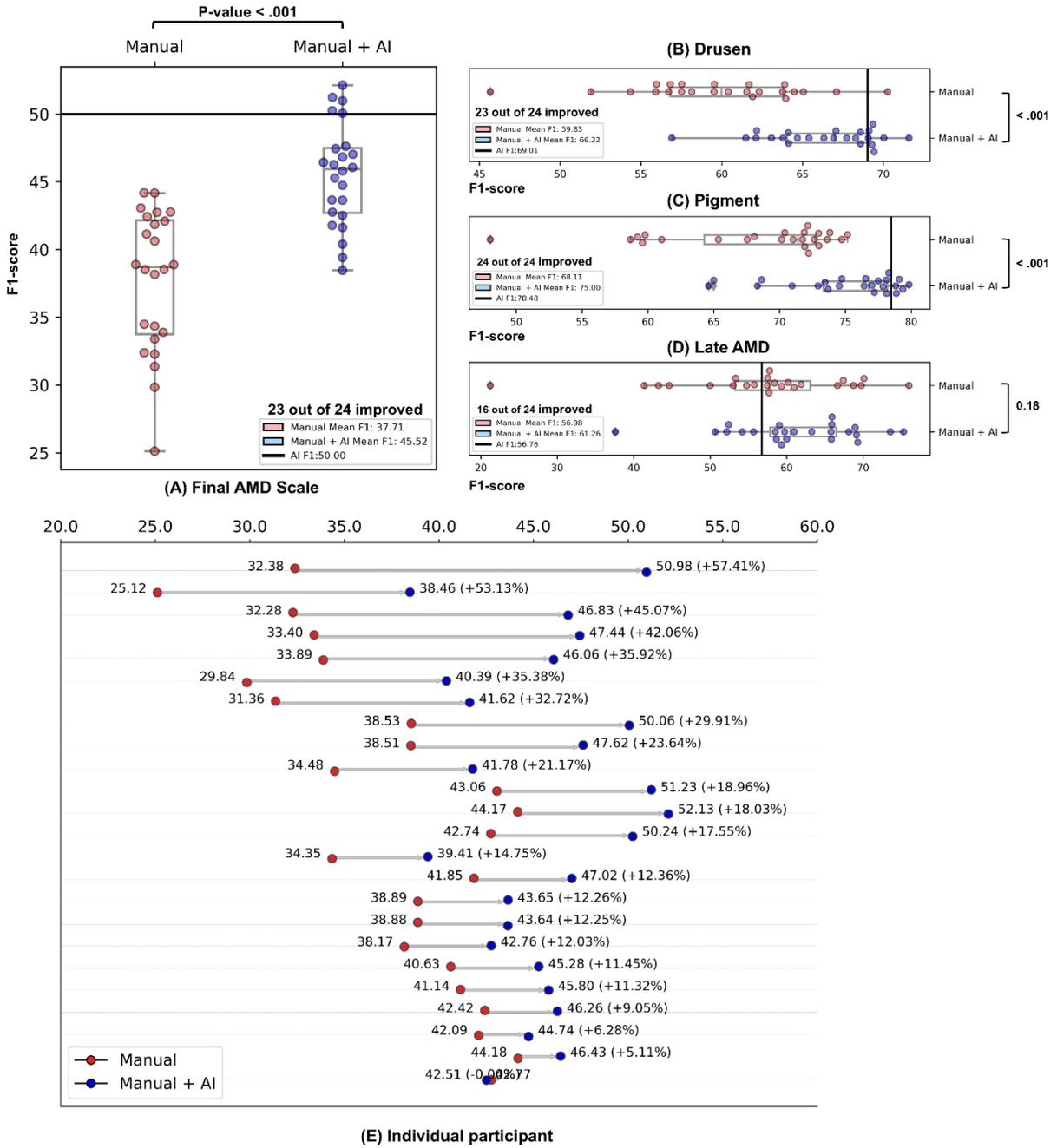

Figure 2. Comparison of Diagnostic Performance (F1-score): Manual vs. Manual + AI Assessment. Each dot represents an F1-score. (A) Final AMD scale and individual risk factors. (B) Drusen. (C) Pigment. (D) Late AMD. The cutoff line represents the performance of the AI model alone. (E) Changes in performance for individual clinicians in terms of F1-score. Blue dots; manual assessment, red dots; AI-assisted manual assessment, solid lines; retina specialists, dashed lines; comprehensive ophthalmologists.

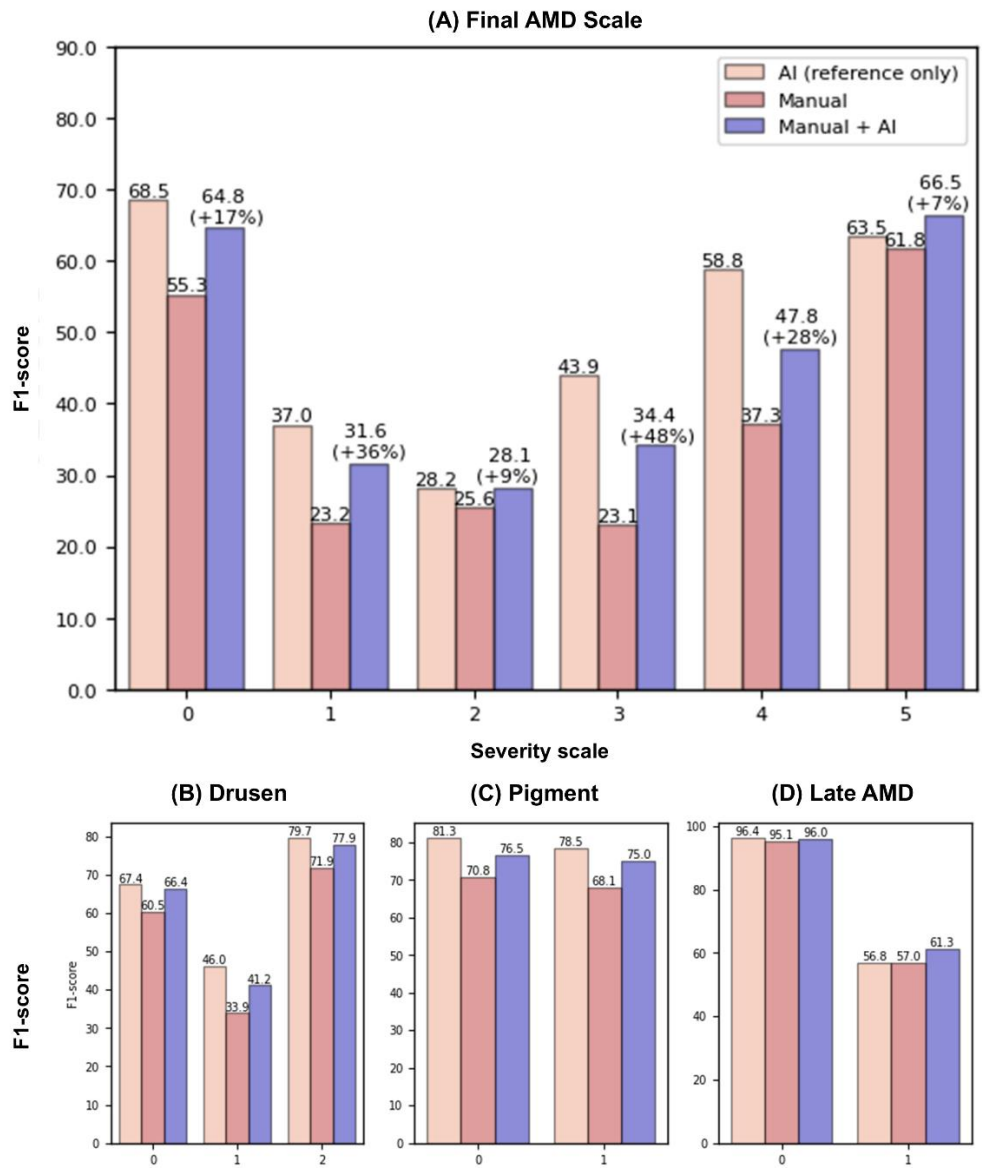

Figure 3. Detailed Breakdown of F1-Score Per Scale for (A) Final AMD Scale, (B) Drusen, (C) Pigment, and (D) Late AMD. Manual and Manual + AI results represent paired comparisons from the same clinicians. The AI-only performance is from a single model and is shown for reference only; it is not directly comparable to the clinician results.

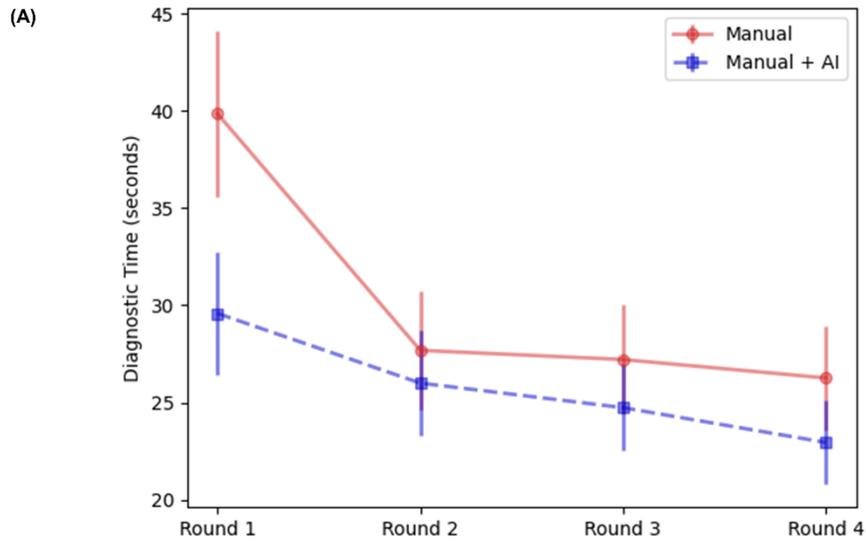
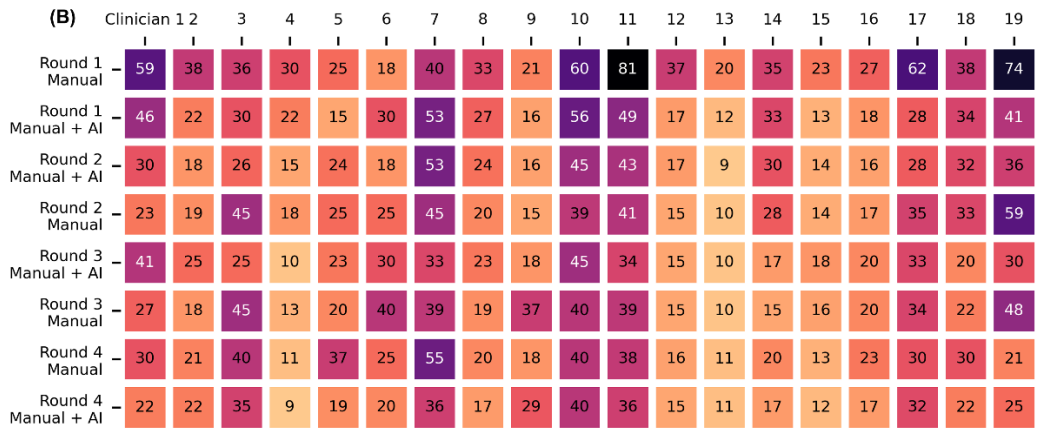

Figure 4. Diagnostic Time (seconds per patient) Efficiency with AI Assistance. (A) Mean and standard deviation of diagnostic times across four rounds. (B) The detailed individual clinician diagnostic time over four rounds, contrasting manual (top row in each round) with AI-assisted assessments (bottom row in each round). Darker shades indicate longer diagnostic times.

Supplementary material

S1 Clinical classification of AMD

S2 Participant details

S3 Training detail

S4 Detailed evaluation results between Manual and Manual + AI

S5 Detailed mixed linear model regression results